\documentclass[12pt]{iopart}

\usepackage{iopams} 
\usepackage{graphicx}  
\begin{document}

\title[]{Inverted random nanopyramids patterning for crystalline silicon photovoltaics}

\author{Ounsi El Daif, Christos Trompoukis, Bjoern Niesen, Marwa Ben Yaala, Parikshit Pratim Sharma, Valerie Depauw, Ivan Gordon}

\address{imec}
\ead{ounsi.eldaif@imec.be}
\begin{abstract}
We demonstrate a nanopatterning technique for silicon photovoltaics, which optically outperforms conventional micron-scale random pyramids, while decreasing by a factor of ten the quantity of silicon lost during the texturing process. We combine hole-mask colloidal lithography, a bottom-up nanolithography technique, with reactive ion etching to define nanopyramids at the surface of a silicon wafer. Thanks to the self-organised aspect of the technique, the beads are randomly distributed, however keeping a interbead distance of the order of their diameter. We tune the nanopattern feature size to maximize the absorption in the crystalline silicon by exploiting both anti-reflection and light trapping. When optimized, the nanopyramids lead to a higher absorption in the crystalline silicon than the conventional micron-scale random pyramids in the visible and near the band edge, with a superior robustness to variations of the angle of the incident light. As the nanopatterning technique presented here is simple, we expect that it could be readily integrated into the crystalline silicon solar cell fabrication processing. 
\end{abstract}

%Uncomment for PACS numbers title message
%\pacs{00.00, 20.00, 42.10}
% Keywords required only for MST, PB, PMB, PM, JOA, JOB? 
%\vspace{2pc}
\noindent{\it Keywords}: Article preparation, IOP journals
% Uncomment for Submitted to journal title message
\submitto{\NT}
% Comment out if separate title page not required
%\maketitle

\section{Introduction}
Since decades, solar cells based on crystalline silicon (cSi) have been dominating the photovoltaic market. Enhancing the power conversion efficiency of this type of solar cell, while, at the same time, decreasing fabrication costs is an important short-term strategy to increase the competiveness of photovoltaic energy generation with respect to conventional energy sources. As the silicon material itself represents an important contribution to the cSi solar cells fabrication costs, the thickness of the cells is one of the issues to tackle. It is necessary to thin down the solar cells as much as possible without compromising their efficiency, and even enhancing it. As cSi has a relatively low absorption coefficient on a broad spectrum, keeping a high efficiency means tackling the losses in absorption, and therefore also in current. In present commercial cSi solar cells, this is commonly done by texturing the front surface of the solar cell with randomly distributed pyramids. These typically have a height of a few microns and a broad size distribution. As this texturing is obtained by anisotropic hydroxide solution etching, the pyramids feature sidewalls with a slope of 54.7° \cite{greenbook}. Record efficiencies for single-junction silicon solar cells were obtained by solar cells textured with periodic inverted pyramids using a multi-step method involving photolithography, which is not compatible with industrial solar cell manufacturing. These inverted pyramids also have a typical scale of several microns, thus much larger than the wavelength of the absorbed visible and near infrared light \cite{green25}. Both these types of textures rely on refraction and reflection of light to enhance the absorption, and as a result, also the photocurrent generated by the solar cell. The commonly accepted absorption enhancement limit for these types of textures is that of a randomly nanotextured surface, with a lambertian scattering pattern, i.e. that scatters all incident light with a cosine angular profile. Another common feature of both these textures is a cSi material consumption of 10-15 $\mu m$ thickness from each wafer sides, which means that they can only be applied to silicon wafers that are much thicker than that. 

Recently, it has been proposed to surpass this absorption efficiency by using a wavelength-scale periodic texturing with a feature size and period of a few hundreds of nanometers, comparable to the wavelength of the desired spectral range (i.e. ~300-1170 nm for cSi \cite{mellorluque}). More specifically, the potential of periodically nanotexturing crystalline silicon layers for light absorption enhancement has recently been demonstrated \cite{herman:113107,trompoukisSPIE}. Moreover, as nanoscale texturing can consume a small quantity of material, it can be applied to solar cells that are only a few microns thick. Several preliminary studies on such very thin cSi cells showed that the patterning steps required to realize the nanoscale texture can be integrated into a cSi cell process flow \cite{Meng:11,trompoukisAPL}. Finally, the use of a subwavelength texture with a tapered profile, resulting in a gradually changing refractive index can assist in coupling light into the cell \cite{herman:113107,deparisJAP,Chattopadhyay20101}.

Here, we demonstrate a nanotexture with a submicron material consumption that yields a pyramidal nanopattern with a short-range and disordered pattern. This allows for broader diffraction resonances compared to a periodic nanopattern, without completely losing diffraction, as it would be the case for a purely random texture \cite{deparisJAP,wiersma}. It produces thus a high absorption on the whole spectrum of interest for silicon solar cells (300-1170nm). The nanotexture is obtained by combining hole-mask colloidal lithography (HCL) \cite{HCL}, a bottom-up lithography technique, in combination with reactive ion etching. The HCL technique was originally developed for the fabrication of metal nanoparticles or nanopatterned metals and has been applied, amongst others, for light trapping in PV \cite{ElDaif201258,niesenplasmon}. We demonstrate the optical performance of such a nanoscale near-random texturing on silicon wafers. First, we will describe our experimental approach and the resulting topography, followed by the discussion of its optical properties as well as the robustness to variations of the angle of incidence of light.

\section{experiments}
\begin{figure}[htbp]
	\begin{center}
		\includegraphics[width=0.5\textwidth]{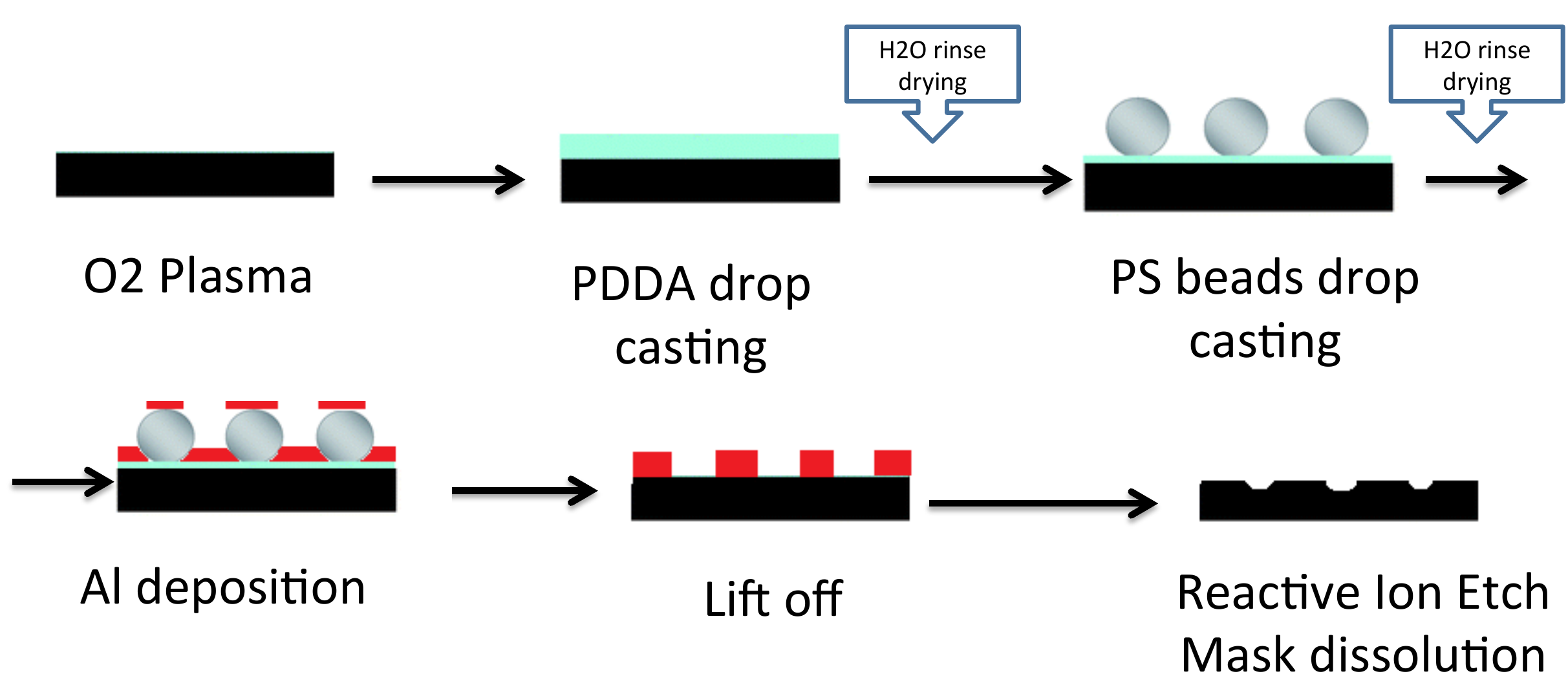}
		\caption{The nanopattern process: hole mask colloidal lithography followed by dry etching.}
		\label{HCLprinciple}
	\end{center}
\end{figure}

Hole-mask colloidal lithography and dry etching were employed to nanopattern the front surface of 700 micron-thick mirror-polished (100) cSi wafers, as schematically shown in Fig. \ref{HCLprinciple}. Samples with a size of 2x2 $cm^{2}$ were solvent-cleaned and treated in an O$_{2}$ plasma for  120 s at 200 W to obtain a hydrophilic surface. This was followed by the subsequent drop casting of an aqueous solution of 0.2 wt\% of poly(diallyldimethylammonium chloride) (PDDA) and a colloidal solution containing 0.2 wt\% negatively charged polystyrene (PS) nanobeads (sulfate latex, invitrogen) with a diameter of either 140, 200, 400, or 500 nm. The sample surface was then rinsed with deionized water, and nitrogen dried. The electrostatic repulsion between the deposited beads leads to a mean surface-to-surface distance similar to their diameter, resulting in a near short-range order. At the same time, the attraction between the beads and the positively charged PDDA allows to keep the particles at a fixed position during the lithography process. A thin film of 50 nm of Al is then deposited by thermal evaporation. Using a residue-free adhesive tape, the beads and the Al covering them are removed, leaving nanoholes in the Al thin film. The spacing between the holes and their diameters are determined by the separation distance and size of the removed nanobeads. The pattern is transfered into the silicon surface with a reactive ion etching (RIE) step using a combination of SF$_{6}$ and O$_{2}$ gases at low pressure. Finally, the Al mask (and possible resist residues) were removed by dissolution in OPD (Tetramethylammonium Hydroxide in water, an organic base used as a positive resist developer due to its reactive properties with metals).  A low concentration is used, i.e. $< 5$ wt\%, in order not to etch silicon as well.

Micron-scale random pyramid textures (typical texturing for industrial wafer based solar cell technologies \cite{greenbook}) were obtained by etching a cSi wafer  in a potassium hydroxide (KOH) solution. They were used for optically benchmarking the nanopattern described above. The resulting microtexturing consists of randomly distributed pyramids with dimensions of 5-7 $\mu m m$ and 15-20 $\mu m m$ material waste during their formation.

The topography of the nanopattern was analyzed utilizing scanning electron microscopy top views (SEM) and cross-section views (XSEM), and its optical properties were assessed with spectrally resolved reflection and absorption measurements using an integrating sphere. The absorption was measured by suspending the sample inside the integrating sphere and measuring at once the reflected and the transmitted light. During these measurements,  any light emerging from the sample is collected. Moreover, this technique offers the possibility of varying the angle of the incident light by rotating the sample inside the sphere.

\section{Results and discussion}
\subsection{Nanopattern topography}

An essential parameter of the nanopatterning method described here is the distribution of the PS beads on the surface of the wafer. It is this distribution that will determine the order and influence the feature size of the pattern transferred into the cSi material in the subsequent RIE etching step. 

\begin{figure}[htbp]
	\begin{center}
		\includegraphics[width=0.5\textwidth]{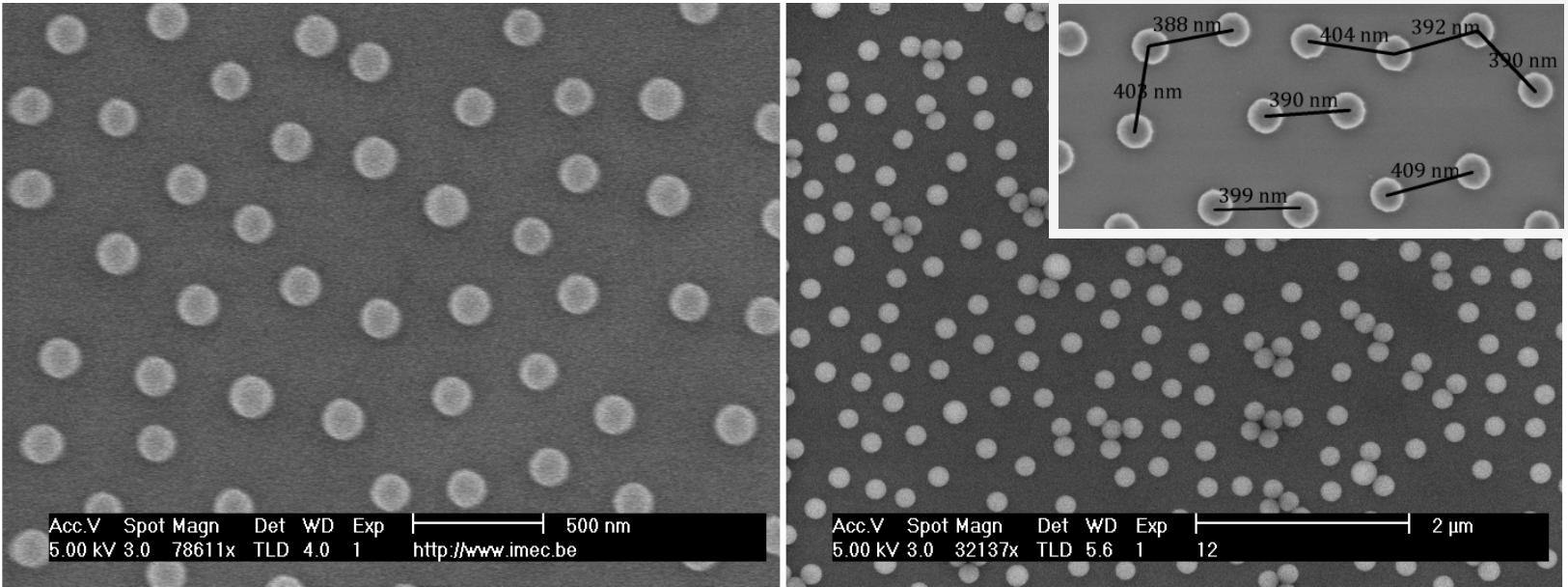}
		\caption{Scanning electron microscope (SEM) imag of the beads.}
		\label{sembeads}
	\end{center}
\end{figure}

As shown in Fig. \ref{sembeads}, the beads have a short-range order with an average border-to-border distance similar to their diameter. The order is due to a combination of effects that include the electrostatic repulsion between the negatively charged PS beads and attraction between the beads and the positively charged wafer surface, as detailed above. In addition, the beads and the liquid strongly interact as the PS bead dispersion dries.

Preserving the colloidal particles at their original position obtained in the initial adsorption process is necessary to obtain a homogeneous well-covered surface. During the drying process, the PS beads are at a given point only partially immersed in water with a subsequent deformation of the liquid surface that gives rise to capillary attractive forces between the beads. These lateral interactions caused by the curved water meniscus between the beads bring about a high tendency for agglomeration. 

It can be shown that in case of partially immersed spherical particles with equal radii the capillary forces are proportional to the square of the radius of the involved particles and to the surface tension of the liquid. Decreasing the surface tension of the liquid can therefore reduce agglomeration. This can be achieved by pre-coating the surface with a non-ionic surfactant (here, Triton X100) or by rinsing it with hot water at  95$^{o}C$ after the beads deposition, since surface tension decreases with temperature.

\begin{figure}[htbp]
	\begin{center}
		\includegraphics[width=0.5\textwidth]{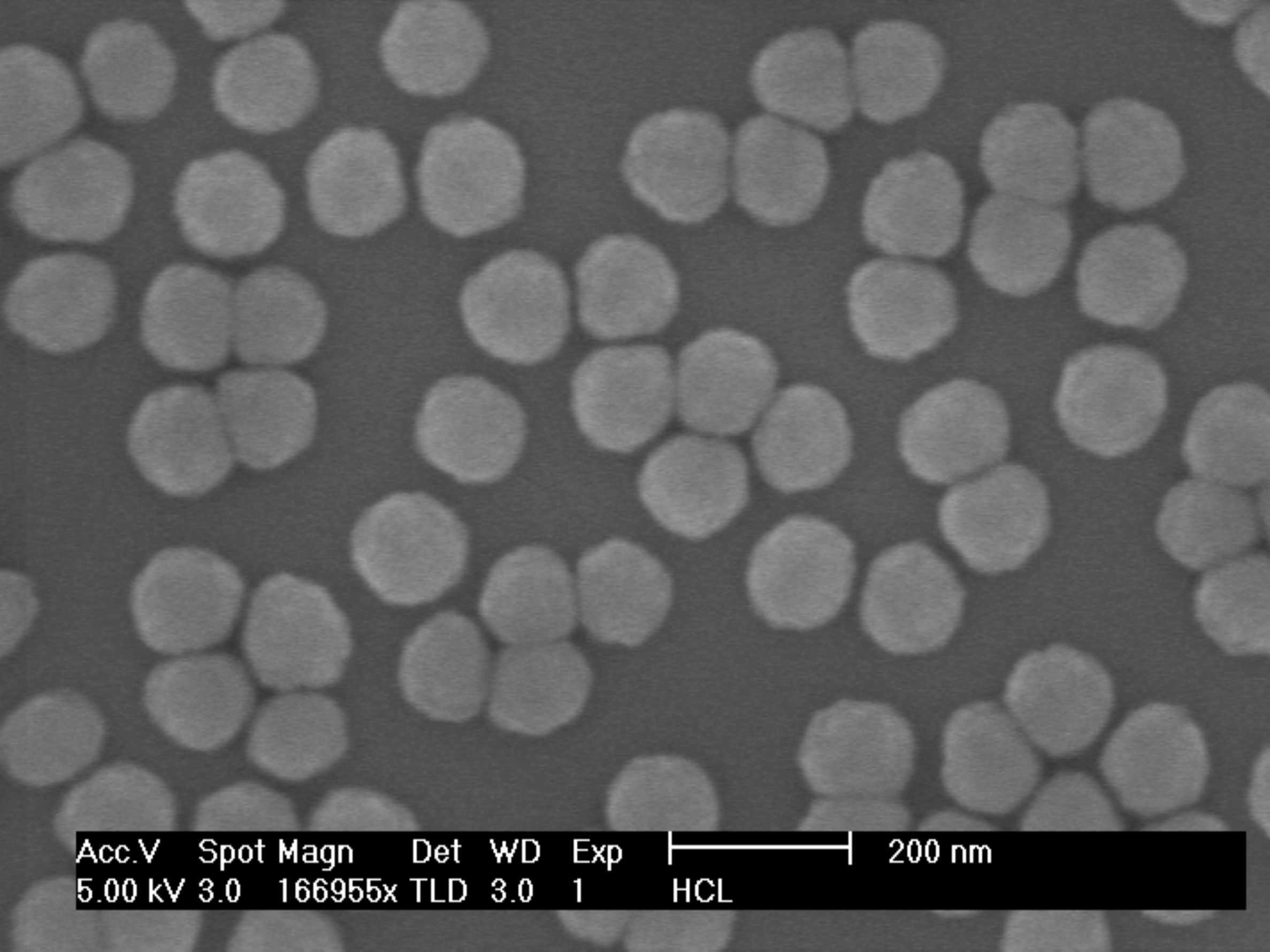}
		\caption{Aggregates when not using surfactant and heating.}
		\label{aggreg}
	\end{center}
\end{figure}

Our standard protocol (schematically shown on Fig. \ref{HCLprinciple}) leads to aggregates (as shown on Fig. \ref{aggreg}) whereas the use of the surfactant and the hot water process further improves the surface distribution and coverage by reducing beads agglomeration (Fig. \ref{sembeads}). However, the square-law dependence of the capillary interaction with the radii of the beads becomes critically high when bigger beads are involved, leading to non-uniform distribution and partial agglomeration.

After the bead deposition, the Si wafer with the PS beads on its surface is coated with a thin film of Al by thermal evaporation, followed by the removal of the beads. This Al layer, that features nanoholes defined by the PS beads, is then employed as an etch mask when the nanopattern is translated into the cSi material by RIE etching.

\begin{figure}[htbp]
	\begin{center}
		\includegraphics[width=0.5\textwidth]{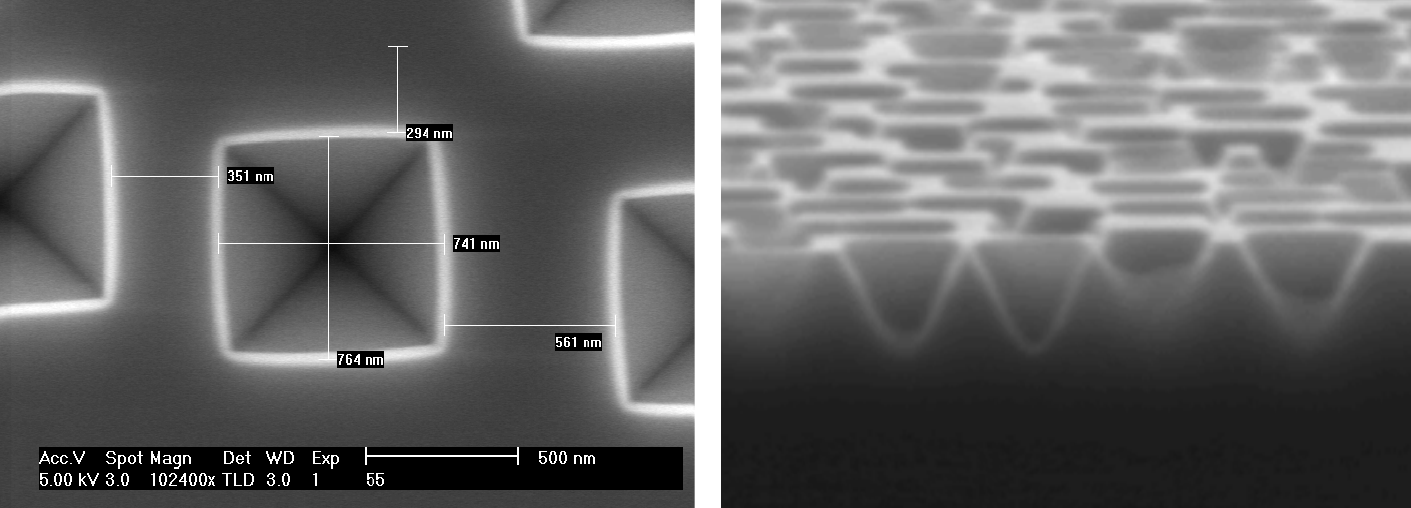}
		\caption{Etched silicon wafer (a) Square shape seen from above (b) Inverted pyramids seen in cross section.}
		\label{shape}
	\end{center}
\end{figure}

As one can see from Fig.\ref{shape}, the shape of a single feature of the pattern is square as seen from above and inverted pyramidal as seen from the side. This remarkable etch shape is due to the mixture of SF$_{6}$ and O$_{2}$ gases during the RIE plasma process step, along with the presence of the Al hard mask, which altogether interact and result in a crystal orientation dependent etching, a similar mechanism has been reported in \cite{tanaka} and we study it in detail in other contributions \cite{PESM,christosetch}.

\subsection{Optical results and comment on the effect of some topographical parameters and the reflectance curves}

For our study, we chose polished crystalline silicon wafers with a thickness of 700 microns. As a result, light with wavelengths of up to $~$1000 nm that is coupled into the wafer will be completely absorbed before it reaches its rear side. However, due to the decreasing absorption coefficient near the cSi bandgap, light in the spectral range between 1000-1170 nm might still reach the rear surface, where it can be reflected. Therefore, at short and average wavelengths, the optical properties of the sample will be completely determined by front side features, whereas light trapping effects by reflection at the rear surface become important for wavelengths near the bandgap.

We performed reflectance measurements utilizing an integrating sphere to determine the optical properties of nanopatterned Si wafer resulting from various PS bead sizes. The integrating sphere allows to measure the full reflectance of the sample on the half-space of incidence, with a near-normal incident angle of 8$^{o}$. 

\begin{figure}[htbp]
	\begin{center}
		\includegraphics[width=0.5\textwidth]{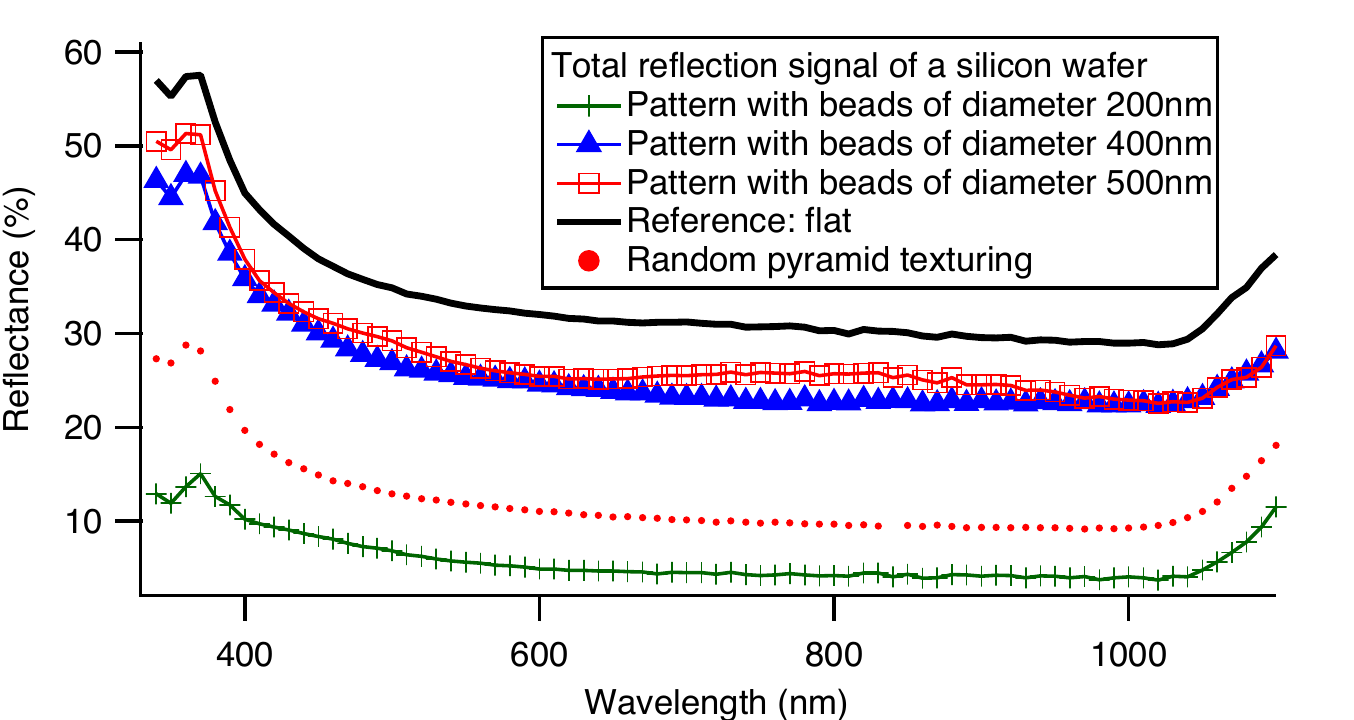}
		\caption{Total reflectance for various bead sizes. The reflectance of a flat silicon substrate and of a traditional micron scale pyramidal pattern are shown for comparison.}
		\label{totrefl}
	\end{center}
\end{figure}

When varying the PS bead size and measuring the reflectance of the nanopatterned Si wafer, we obtain spectra with similar spectral shape, but with very different absolute values. One can see on Figure \ref{totrefl} that the total reflectance curves vary from a level similar to that of bare silicon to a very low level, in average below 10\%. The highest reflectance is given by the largest beads. This difference can be explained both by the agglomeration of big beads during the HCL process (see Figure \ref{aggreg} above) as well as the increasing average bead-to-bead distance. This leads to large fractions of the wafer surface that remain flat for the larger PS bead sizes, reflecting a significant amount of light. As a comparison, the total reflectance of a flat polished and of a pyramid microtextured cSi wafers are shown. The reflectance given by the HCL nanopatterning is around 5\%, below the traditional pyramidal patterning \cite{greenbook} shown in red dots on the figure. 

In addition to reflectance measurements we performed direct absorption measurements to further compare the properties of this nanopatterning to the state of the art used in typical cSi solar cells, which contains random pyramids of 5-10 microns size, fabricated by wet etching in KOH. We directly measured the sum of reflection and transmission together by suspending the sample inside the sphere, and deduced the absorption as A=1-R-T. This allowed us to simultaneously vary the angle of incidence, in order to study the angular robustness of our results. 

\begin{figure}[htbp]
	\begin{center}
		\includegraphics[width=0.5\textwidth]{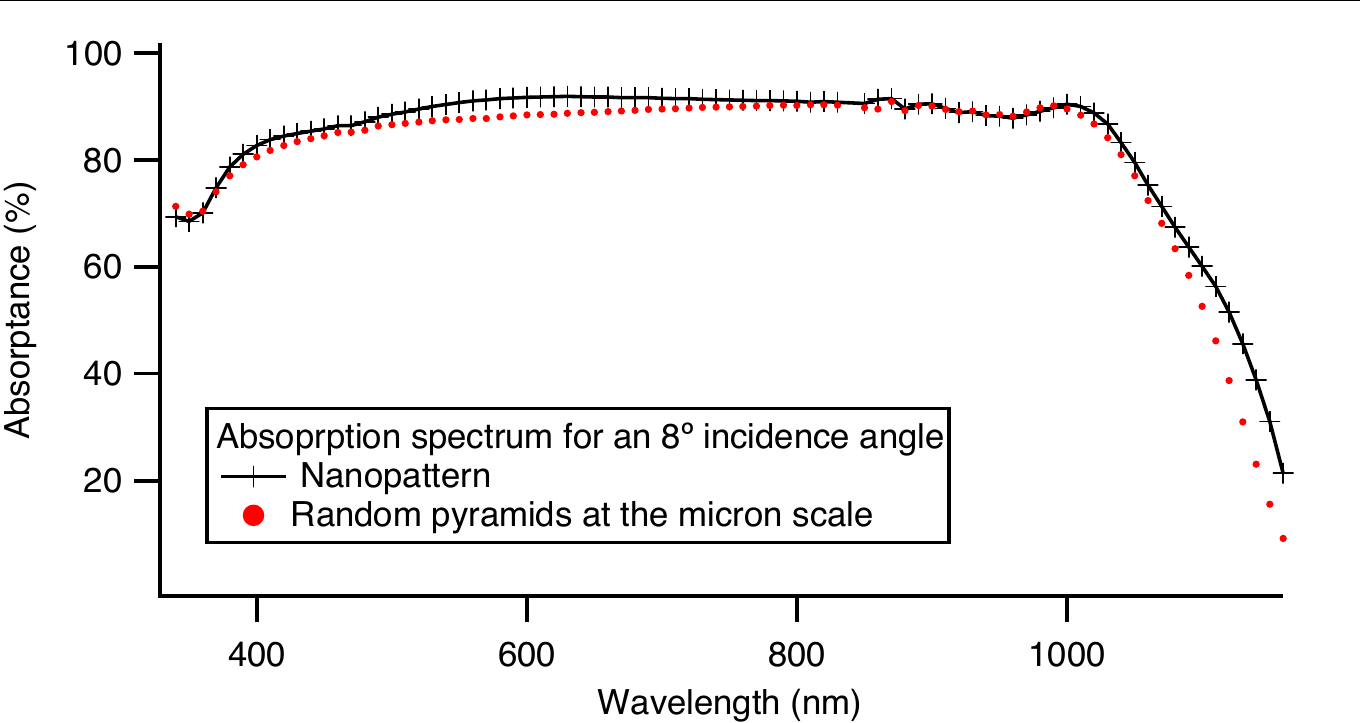}
		\caption{Absorption spectra for the best nanopatterning as well as for random pyramid textured silicon}
		\label{abs}
	\end{center}
\end{figure}

One can see on Fig.\ref{abs} that the nanopattern yields an absorption in the cSi wafer which is mostly surpassing that of the random pyramids. Angularly, the achieved result is even better. We explain this combination of results by the following aspects: 
\begin{itemize}
\item the front side reflectance is decreased thanks to the pyramidal shape of the front side nanopatterning by a progressive change in the refractive index. It has been shown \cite{herman:113107} that at dimensions of patterning equal or smaller than the wavelength range, non-vertical sidewalls allow an effect similar to a graded index effect, and help further decrease the reflectivity compared to perfect vertical sidewalls.
\item Concerning the absorption at long wavelength, i.e. near the band edge, although a periodic patterning at the wavelength scale is expected to enhance the light trapping for selected wavelengths \cite{mellorluque,wiersma}, it has been demonstrated for the case of thin membranes, that a dose of disorder can help broadening the resonances offered by perfect periodicity. In fact Vynck et al. \cite{wiersma} propose a novel concept of amorphous photonic quasicrystal where they show that, on one hand adding disorder in the photonic lattice decreases slightly the photon lifetime of the structure because the modes quality factor decreases, and on the other hand, for the same reasons, a symmetric behavior of resonance spectral broadening appears. We do not deal here with a membrane but with a relatively thick layer compared to the optical wavelength, however, a front side periodic nanopatterning would still show diffraction resonances, on which a slight disorder may have a similar effect of resonance broadening.
\end{itemize}

\begin{figure}[htbp]
	\begin{center}
		\includegraphics[width=0.5\textwidth]{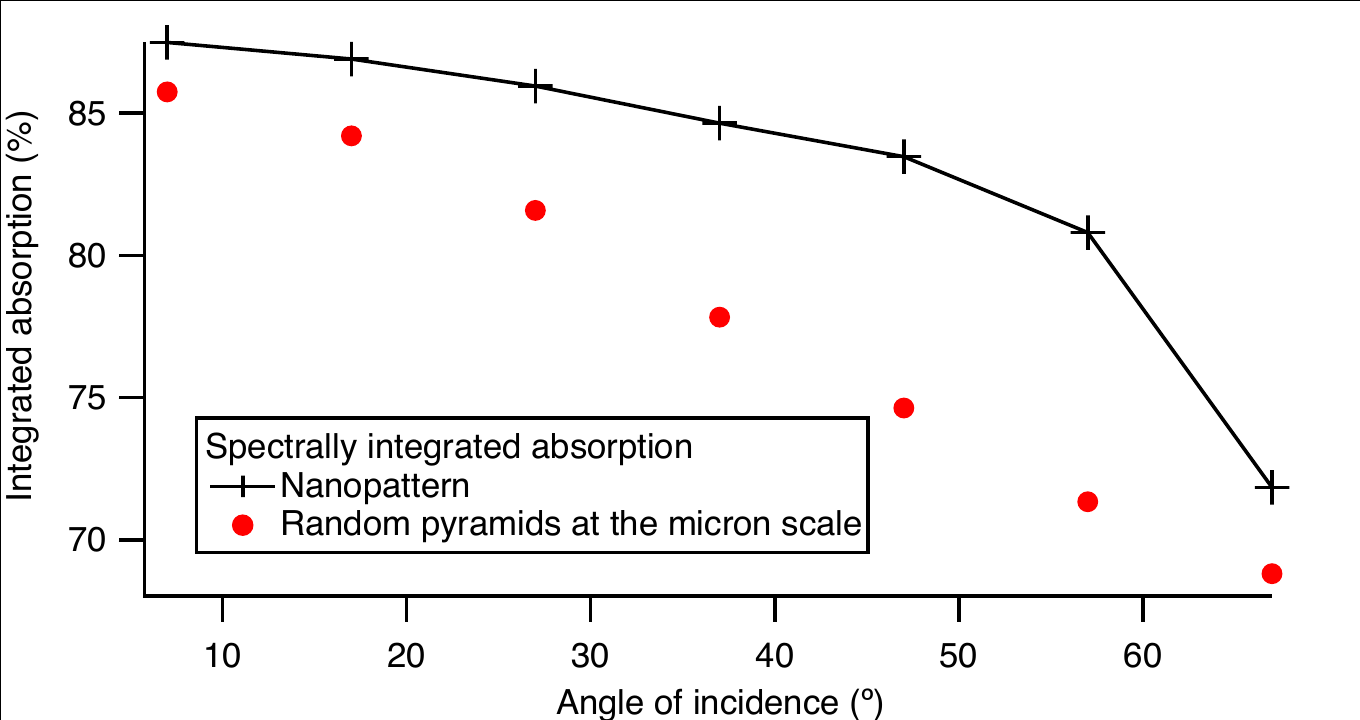}
		\caption{Evolution of the integrated absorption of the two patternings (with an AM15 standard solar spectrum) with respect to the incidence angle of light}
		\label{angle}
	\end{center}
\end{figure}

Finally the angular evolution of the absorption is an important aspect for the annual energy yield of a solar cell device. Present solar cells lose around 3\% \cite{4060141} of their annual yield due to the angular mismatch between laboratory and real illumination conditions. Therefore, a better angular robustness is an important added value and contributes to increasing the solar modules yield, in addition to increasing the AM 1.5G efficiency. As opposed to micropatterning, where the geometrical aspect of the patterning yields a very strong dependence on the angle of incidence because the incident light interacts with flat surfaces at fixed angles, nanopatterning acts on the wave nature of light, and increasing the angle of incidence of light has mainly the impact of slightly changing (by a cosine factor) the effective dimension of the pattern seen by the incident light. This can slightly shift resonances, but given the broad character of the resonance we are dealing with, the effect on the properties on a broad spectrum is negligible. This results in a very robust angular behavior.

\section{Conclusion}

We present a nanopattern that efficiently acts as an anti-reflection and light trapping layer when applied at the front side of cSi wafers. The absorption enhancement achieved with this nanopattern is higher than that of random micron-sized pyramids, while consuming less than one micron thickness of material. This reduced material consumption is currently gaining interest due to present developments in decreasing silicon consumption and therefore cSi solar cell thickness to reduce costs. The recent development in thin crystalline silicon foils (one to forty micron) processing \cite{drosspip} can therefore fully benefit from this patterning. The cSi nanopattern presented here with its excellent antireflection and light trapping capabilities might therefore find application in highly efficient, ultra-thin cSi solar cells. \cite{petermanPIP,depauwPIP,drosspip}.

Alternatively despite the importance of decreasing the material consumption, the PV system cost is a difficult item to decrease. This means that the efficiency of solar cells, even thinner thanks to the novel technology presented here, needs to stay high. Therefore, given the proof-of-concept presented here, an optimization of the nanopatterning parameters is the pathway to even further increasing the maximum current achievable.

\section*{Acknoweledgements}

This work has partially been performed in the frame of the PhotoNVoltaics FP7 European project, grant number 309127 and the PRIMA European project, grant number 248154. We would like to thank Prof. Olivier Deparis and Aline Herman from the University of Namur, as well as Prof. A. Dmitriev, Dr L. Tong and Dr. Vladimir Miljkovic from Chalmers for fruitful discussions on metal nanoparticles fabrication.

\section*{References}

\bibliographystyle{unsrt} %cf le doc biblio-pdf pour des commandes marrantes sur la biblio
\bibliography{biblio}

\begin{thebibliography}{10}

\bibitem{greenbook}
Marten Green.
\newblock {\em Silicon solar cells advanced principles and practice}.
\newblock Center for photovoltaic devices and systems, Sydney, 1995.

\bibitem{green25}
Martin~A. Green.
\newblock The path to 25% silicon solar cell efficiency: History of silicon
  cell evolution.
\newblock {\em PROGRESS IN PHOTOVOLTAICS: RESEARCH AND APPLICATIONS},
  17:183--189, 2009.

\bibitem{mellorluque}
A.~Marti M.J.~Mendes A.~Mellor, I.~Tobias and A.~Luque.
\newblock Upper limits to absorption enhancement in thick solar upper limits to
  absorption enhancement in thick solar cells using diffraction gratings.
\newblock {\em PROGRESS IN PHOTOVOLTAICS: RESEARCH AND APPLICATIONS},
  19:676--687, 2011.

\bibitem{herman:113107}
Aline Herman, Christos Trompoukis, Valerie Depauw, Ounsi~El Daif, and Olivier
  Deparis.
\newblock Influence of the pattern shape on the efficiency of front-side
  periodically patterned ultrathin crystalline silicon solar cells.
\newblock {\em Journal of Applied Physics}, 112(11):113107, 2012.

\bibitem{trompoukisSPIE}
Christos Trompoukis, Aline Herman, Ounsi El~Daif, Valerie Depauw, Dries
  Van~Gestel, Kris Van~Nieuwenhuysen, Ivan Gordon, Olivier Deparis, and Jef
  Poortmans.
\newblock Enhanced absorption in thin crystalline silicon films for solar cells
  by nanoimprint lithography.
\newblock pages 84380R--84380R--10, 2012.

\bibitem{Meng:11}
Xianqin Meng, Valerie Depauw, Guillaume Gomard, Ounsi~El Daif, Christos
  Trompoukis, Emmanuel Drouard, Alain Fave, Frederic Dross, Ivan Gordon, and
  Christian Seassal.
\newblock Design and fabrication of photonic crystals in epitaxial free silicon
  for ultrathin solar cells.
\newblock In {\em Display, Solid-State Lighting, Photovoltaics, and
  Optoelectronics in Energy}, page 831207. Optical Society of America, 2011.

\bibitem{trompoukisAPL}
Christos Trompoukis, Ounsi~El Daif, Valerie Depauw, Ivan Gordon, and Jef
  Poortmans.
\newblock Photonic assisted light trapping integrated in ultrathin crystalline
  silicon solar cells by nanoimprint lithography.
\newblock {\em Applied Physics Letters}, 101(10):103901, 2012.

\bibitem{deparisJAP}
Olivier Deparis, Jean~Pol Vigneron, Otto Agustsson, and Daniel Decroupet.
\newblock Optimization of photonics for corrugated thin-film solar cells.
\newblock {\em Journal of Applied Physics}, 106(9):094505, 2009.

\bibitem{Chattopadhyay20101}
S.~Chattopadhyay, Y.F. Huang, Y.J. Jen, A.~Ganguly, K.H. Chen, and L.C. Chen.
\newblock Anti-reflecting and photonic nanostructures.
\newblock {\em Materials Science and Engineering: R: Reports}, 69(1--3):1 --
  35, 2010.

\bibitem{wiersma}
Francesco~Riboli Kevin~Vynck, Matteo~Burresi and Diederik~S. Wiersma.
\newblock Photon management in two-dimensional disordered media.
\newblock {\em Nature Materials}, 11:1017--1022, 2012.

\bibitem{HCL}
H.~Fredriksson, Y.~Alaverdyan, A.~Dmitriev, C.~Langhammer, D.~S. Sutherland,
  M.~Z{\"a}ch, and B.~Kasemo.
\newblock Hole--mask colloidal lithography.
\newblock {\em Advanced Materials}, 19(23):4297--4302, 2007.

\bibitem{ElDaif201258}
Ounsi~El Daif, Lianming Tong, Bruno Figeys, Kris~Van Nieuwenhuysen, Alexander
  Dmitriev, Pol~Van Dorpe, Ivan Gordon, and Frederic Dross.
\newblock Front side plasmonic effect on thin silicon epitaxial solar cells.
\newblock {\em Solar Energy Materials and Solar Cells}, 104(0):58 -- 63, 2012.

\bibitem{niesenplasmon}
Bjoern Niesen, Barry~P. Rand, Pol Van~Dorpe, David Cheyns, Lianming Tong,
  Alexandre Dmitriev, and Paul Heremans.
\newblock Plasmonic efficiency enhancement of high performance organic solar
  cells with a nanostructured rear electrode.
\newblock {\em Advanced Energy Materials}, 3(2):145--150, 2013.

\bibitem{tanaka}
S.~Tanaka, K.~Sonoda, K.~Kasai, K.~Kanda, T.~Fujita, K.~Higuchi, and
  K.~Maenaka.
\newblock Crystal orientation dependent etching in rie and its application.
\newblock In {\em Micro Electro Mechanical Systems (MEMS), 2011 IEEE 24th
  International Conference on}, pages 217--220, 2011.

\bibitem{PESM}
Christos Trompoukis.
\newblock Crystal orientation dependent anisotropic dry silicon etching.
\newblock In {\em PESM Conference (Leuven, Belgium)}, 2013.

\bibitem{christosetch}
Christos Trompoukis, Ounsi El~Daif, Valerie Depauw, Parikshit Sharma, and Jef
  Poortmans.
\newblock Crystal orientation dependent anisotropic dry silicon etching.
\newblock Unpublished, 2013.

\bibitem{4060141}
T.~Oozeki, K.~Otani, and K.~Kurokawa.
\newblock An evaluation method for pv system to identify system losses by means
  of utilizing monitoring data.
\newblock In {\em Photovoltaic Energy Conversion, Conference Record of the 2006
  IEEE 4th World Conference on}, volume~2, pages 2319--2322, 2006.

\bibitem{drosspip}
Frederic Dross, Kris Baert, Twan Bearda, Jan Deckers, Valerie Depauw, Ounsi
  El~Daif, Ivan Gordon, Adel Gougam, Jonathan Govaerts, Stefano Granata, Riet
  Labie, Xavier Loozen, Roberto Martini, Alex Masolin, Barry O'Sullivan,
  Yu~Qiu, Jan Vaes, Dries Van~Gestel, Jan Van~Hoeymissen, Anja Vanleenhove,
  Kris Van~Nieuwenhuysen, Srisaran Venkatachalam, Marc Meuris, and Jef
  Poortmans.
\newblock Crystalline thin-foil silicon solar cells: where crystalline quality
  meets thin-film processing.
\newblock {\em Progress in Photovoltaics: Research and Applications},
  20(6):770--784, 2012.

\bibitem{petermanPIP}
Jan~Hendrik Petermann, Dimitri Zielke, Jan Schmidt, Felix Haase,
  Enrique~Garralaga Rojas, and Rolf Brendel.
\newblock 19%-efficient and 43 µm-thick crystalline si solar cell from layer
  transfer using porous silicon.
\newblock {\em Progress in Photovoltaics: Research and Applications},
  20(1):1--5, 2012.

\bibitem{depauwPIP}
V.~Depauw, Y.~Qiu, K.~Van~Nieuwenhuysen, I.~Gordon, and J.~Poortmans.
\newblock Epitaxy-free monocrystalline silicon thin film: first steps beyond
  proof-of-concept solar cells.
\newblock {\em Progress in Photovoltaics: Research and Applications},
  19(7):844--850, 2011.

\end{thebibliography}

\end{document}